\documentclass[12pt]{article}
\usepackage{epsf}
\usepackage{graphicx}

\begin{document}
\title
{A few remarks about the Pioneer anomaly}
\author
{Michael A. Ivanov \\
Physics Dept.,\\
Belarus State University of Informatics and Radioelectronics, \\
6 P. Brovka Street,  BY 220027, Minsk, Republic of Belarus.\\
E-mail: michai@mail.by.}

\maketitle

\begin{abstract} Some features of the Pioneer anomaly are discussed
in context of author's explanation of this effect as a
deceleration of the probes in the graviton background. It is noted
that if the model is true then the best parameter of the anomalous
acceleration should be not the distance to the Sun but a cosine of
the angle between a velocity of the probe and its radius-vector.
\end{abstract}

As it was reported by the authors of the discovery, NASA
deep-space probes Pioneer 10/11 experience an anomalous constant
acceleration directed towards the Sun (the Pioneer anomaly)
\cite{1,2}. A possible origin of the effect remains unknown. In my
model \cite{5}, any massive body must experience a constant
deceleration $w \simeq - Hc$, where $H$ is the Hubble constant and
$c$ is the light velocity, of the same order of magnitude as
observed for cosmic probes. This effect is an analogue of
cosmological redshifts in the model. Their common nature is
forehead collisions with gravitons. I would like to consider here
the main known features of this anomaly in context of my
explanation keeping in mind present and future efforts to verify
the reality of this effect and to understand it.
\par
The observed anomaly has the following main features: 1) in the
range 5 - 15 AU from the Sun it is observed an anomalous sunward
acceleration with the rising modulus which gets its maximum value,
leastwise for Pioneer 11 (see Fig. 3 in \cite{2}); 2) for greater
distances, this maximum sunward acceleration remains almost
constant for both Pioneers \cite{1,2}; 3) also it is observed an
unmodeled annual periodic term in residuals for Pioneer 10
\cite{311} which is obviously connected with the motion of the
Earth.
\par
If my conjecture \cite{5} about the quantum nature of this
acceleration is true then an observed value of the projection of
the probe's acceleration on the sunward direction $w_{s}$ should
depend on accelerations of the probe, the Earth and the Sun
relative to the graviton background. If we assume that the Sun
moves relative to the background slowly enough, then anomalous
accelerations of the Earth and the probe will be directed almost
against their velocities in the heliocentric frame, and in this
case:  $w_{s}=-w\cdot cos\alpha$, where $\alpha$ is an angle
between a radius-vector of the probe and its velocity in the
frame. For a terrestrial observer, an additional term should be
taken into account which is connected with its own motion relative
to the background. \par By the very elongate orbits of the both
Pioneers (see Fig. 3 in \cite{2}), it would explain the second
(and main) peculiarity. For example, for Pioneer 10 at the
distance 67 AU from the Sun one has $\sin\alpha\approx0.11$ (it is
a visual estimate with Fig. 3 of \cite{2}), i.e.
$cos\alpha\approx0.994.$ If for big distances from the Sun we use
the conservation laws of energy and angular momentum in the field
{\it of the Sun only}, then in the range 6.7 - 67 AU a value of
$cos\alpha$ changes from 0.942 to 0.994, i.e. approximately on 5
per cent only. Due to this fact, a projection of the probe's
acceleration on the sunward direction would be almost constant.
\par
As Toth and Turyshev report \cite{3}, they intend to carry out an
analysis of newly recovered data received from Pioneers, with
these data are now available for Pioneer 11 for distances 1.01 -
41.7 AU. If the serious problem of taking into account the solar
radiation pressure at small distances is precisely solved
(modeled) \cite{4}, then this range will be very lucky to confront
the expression $w_{s}=-w\cdot cos\alpha$ of the considered model
with observations for small distances when Pioneer 11 executed its
planetary encounters with Jupiter and Saturn. In this period, a
value of $cos\alpha$ was changed in the non-trivial manner, and
the projection of anomalous acceleration should behave itself
similar. For example, when the spacecraft went to Saturn,
$cos\alpha$ was {\it negative} during some time. If this model is
true, the anomaly in this small period should have {\it the
opposite sign}. I think, it would be the best of all to compare
the two functions of the probe's proper time: the projection of
anomalous acceleration of Pioneer 11 and $cos\alpha$ for it. These
functions should be very similar to each other if my conjecture is
true. At present, a new mission to test the anomaly is planned
\cite{44}. It is seen from this consideration that it would be
desirable to have a closed orbit for this future probe (or the one
with two elongate branches where the probe moves off the Sun and
towards it).
\par
Leaving for the future the question about the stability and form
of the Earth orbit by such the anomalous acceleration, I note that
namely this one would cause feature 3) of the anomaly. In this
case, because Pioneers 10 and 11 have different trajectories, it
is possible to compute a sign of the projection of Earth's
anomalous acceleration contribution: when the Earth moves after a
probe, we should observe a minimum of the periodic term, and we
should see a maximum when they move in opposite directions. For
the twins, these maximums-minimums will appear in different time
intervals, that is important to test the model.
\par
The observed very tiny anomaly in the probe motion may be the
first egress beyond the applicability limits of general relativity
in the solar system. If my explanation of the one is true then
this effect may turn out to be and the first observable
macroscopic manifestation of low-energy quantum gravity.

\end{document}